\documentclass[12pt]{article}
\usepackage{amssymb}
\usepackage{pifont}
\usepackage{mathrsfs}
\usepackage{amsfonts}
\usepackage{color}
\usepackage{footmisc}
\usepackage[numbers,sort&compress]{natbib}
\usepackage[bookmarksopen=true, bookmarksopenlevel=\maxdimen,
bookmarksnumbered=true,colorlinks,linkcolor=blue,
citecolor=blue,urlcolor=blue]{hyperref}

\newcommand{\omits}[1]{}

\begin{document}

\begin{center}
{\bf \LARGE Energy, momentum and angular momentum\\ \bigskip
conservations in de Sitter special relativity}

\bigskip

\bigskip

{\large Jia-An Lu$^{a}$\footnote{Email: ljagdgz@163.com}}

\bigskip

$^a$ School of Physics and Astronomy, Sun Yat-sen University\\
Guangzhou 510275, China \bigskip

\begin{abstract}
In de Sitter (dS) special relativity (SR), two kinds of conserved currents are derived.
The first kind is a 5-dimensional (5d) dS-covariant angular momentum (AM) current, which
unites the energy-momentum (EM) and 4d AM current in an inertial-type coordinate system.
The second kind is a dS-invariant AM current, which can be generalized to a conserved
current for the coupling system of the matter field and gravitational field in dS gravity.
Moreover, an inherent EM tensor is predicted, which comes from the spin part of the
dS-covariant current. All the above results are compared to the ordinary SR with Lorentz invariance.
\end{abstract}
\end{center}



\setcounter{footnote}{-1} \footnote{The final publication is
available at Springer via \href{http://dx.doi.org/10.1007/s10714-015-2001-6}
{http://dx.doi.org/10.1007/s10714-015-2001-6}.}

\section{Introduction}

The de Sitter (dS) special relativity (SR) \cite{Dirac,Gursey,Fantappie,Pessa,Guo-MPL,Guo-PLA,
Yan,Xu,Aldrovandi} is a hypothetical theory with global dS invariance, which is well motivated,
at least for the following reasons. Firstly, the dS symmetry is one of the three highest
symmetries for a 4-dimensional (4d) metric with Lorentz signature \cite{Fantappie,Weinberg}.
Secondly, the observed cosmological constant may be related to that of dS SR, with new observable
effects to be investigated. Thirdly, in dS space, there exist inertial-type coordinate (IC) systems
where all the coordinate lines are geodesics \cite{Pessa,Guo-MPL}; the energy, momentum and 4d
angular momentum (EMA) of a classical particle can be defined in the IC systems, and unified in
a 5d dS-covariant angular momentum (AM) function \cite{Pessa,Guo-MPL}.
Moreover, it is shown that \cite{Lu14,Lu14-b} in the $R+\beta S^{abc}S_{abc}$
models of gravity \cite{Vignolo}, where $R$ is the scalar curvature, $\beta$ is a parameter,
and $S_{abc}$ denotes the torsion tensor, the dS symmetry together with a Kaluza--Klein-type ansatz
may play a key role in removing the initial singularity of the Robertson--Walker universe.

In this paper, we are going to find the definitions of the EMA currents for a
dS-covariant matter field in dS SR, which should be important for the foundation of dS SR
as well as the quantum field theory on dS space \cite{Dirac,Gursey,Xu,Bros,Banerjee,Cortez}.
To do this, we study the conservation law corresponding to the dS and
diffeomorphism symmetries of the matter field. It is shown that, the conserved current is a
5d dS-covariant AM current, whose components can be
identified as the EMA currents in an IC system. Although
the energy-momentum (EM) current is not covariant, the EM tensor can be well defined,
and it is found that the spin part of the dS-covariant current contributes to the EM tensor,
with the contribution proportional to $\Lambda^{1/2}$, where $\Lambda$ is the cosmological constant.

Note that the generalization of the dS-covariant current in dS gravity
\cite{Stelle,PoI,Guo07,Lu13,Lu14} is a variational derivative with respect to
the dS connection \cite{Lu15}. For the coupling system of matter and gravity,
the variational derivative should be equal to zero on account of the gravitational
field equation. Accordingly, we also analyze another kind of conservation law
in dS SR, with respect to
a one-parameter group of entangled dS transformations, which consist of the internal
dS rotations and external isometric transformations, sharing the same generators.
The corresponding conserved current is a 5d dS-invariant AM current, which can be
generalized to a conserved current for the coupling system of the matter field and
gravitational field.

The paper is arranged as follows. In section 2, the EMA conservation laws are reviewed
in the ordinary SR with Lorentz invariance. In section 3, the EMA
conservation laws are derived in dS SR, and it is shown that the EM tensor contains
an inherent part, in addition to the ordinary orbital part.
Finally we end with some remarks in the last section.

\section{Lorentz special relativity}

In the ordinary SR on the Minkowski space ${\cal M}_m$, consider a Lorentz-covariant matter field $\psi$
with the action integral $S$ and Lagrangian function $\mathscr{L}$ as follows:
\begin{equation}
S=\int_\Omega\mathscr{L}\epsilon,\ \
\mathscr{L}=\mathscr{L}(\psi,\partial_a\psi,c.c., e^\alpha{}_a),
\end{equation}
where $\Omega$ is an arbitrary domain of ${\cal M}_m$,
$\epsilon$ is a metric-compatible volume element, $\partial_a$ is the ordinary
derivative of any Minkowski coordinate system, $a$ is an abstract
index \cite{Wald,Liang} which can be changed into any tetrad or coordinate index
by taking the corresponding component, $c.c.$ denotes the complex conjugates of the
former quantities, and $e^\alpha{}_a$ is the dual of an orthonormal tetrad field
$e_\alpha{}^a$, with the tetrad index $\alpha=0,1,2,3$. Let
\begin{equation}\label{symtran}
\Omega\rightarrow\Omega_t=\phi_t[\Omega],\ \
\psi\rightarrow \psi_t=T(g_t)\phi_{t\ast}\psi,\ \
e^\alpha{}_a\rightarrow e_t^\alpha{}_a=g_t^\alpha{}_\beta\phi_{t\ast}e^\beta{}_a
\end{equation}
be symmetry transformations forming a one-parameter group, where $t$ is the group parameter,
$g_t=g_t^\alpha{}_\beta$ are elements of the special Lorentz group $SO(1,3)$,
$T$ stands for a representation of $O(1,3)$, $\phi_t$ are diffeomorphisms,
and $\phi_{t\ast}$ denote their pushforwards. Suppose that the matter field equation
$\delta S/\delta\psi=0$ is satisfied, it follows from the
invariance of the action integral under Eq. (\ref{symtran}) that
\begin{equation}\label{dJ}
\partial_aJ^a+\frac{\partial\mathscr{L}}{\partial e^\alpha{}_a}\delta e^\alpha{}_a
=\mathscr{L}\partial_a v^a,
\end{equation}
where
\begin{equation}\label{J}
J^a=\frac{\partial\mathscr{L}}{\partial\partial_a\psi}\delta\psi
+c.c.+\mathscr{L}v^a,
\end{equation}
$\delta=(d/dt)|_{t=0}$, and $v^a$ is the generator of $\phi_t$.
Making use of the arbitrariness of $A^\alpha{}_\beta\equiv\delta g^\alpha{}_\beta$,
$v^a$ and $\partial_av^b$ at any given point,
it follows that
\begin{equation}\label{dsigma=0}
\partial_a\Sigma_b{}^a=0,
\end{equation}
\begin{equation}\label{dtau=sigma}
\partial_a\tau_{\alpha\beta}{}^a=-\Sigma_{[\alpha\beta]},
\end{equation}
\begin{equation}
-\frac{\partial\mathscr{L}}{\partial d_a\psi}d_b\psi+c.c.
=\frac{\partial\mathscr{L}}{\partial e^\alpha{}_a}e^\alpha{}_b,
\end{equation}
where
\begin{equation}\label{sigma}
\Sigma_b{}^a=-\frac{\partial\mathscr{L}}{\partial\partial_a\psi}\partial_b\psi
+c.c.+\mathscr{L}\delta^a{}_b
\end{equation}
is the canonical EM tensor,
\begin{equation}\label{tau}
\tau_\alpha{}^{\beta a}=\frac{\partial\mathscr{L}}{\partial\partial_a\psi}T_\alpha{}^\beta\psi+c.c.
\end{equation}
is the Lorentz-covariant spin current, $T_\alpha{}^\beta$ are representations of the Lorentz generators,
and the tetrad indices are lowered by $\eta_{\alpha\beta}={\rm diag}(-1,1,1,1)$.
According to Eqs. (\ref{dsigma=0})--(\ref{dtau=sigma}),
$\partial_aJ_\alpha{}^a=\partial_aJ_\alpha{}^{\beta a}=0$ (see, for instance, Ref. \cite{Peskin}), where
\begin{equation}\label{EM}
J_\alpha{}^a=\Sigma_\alpha{}^a,
\end{equation}
\begin{equation}\label{AM}
J_\alpha{}^{\beta a}=\tau_\alpha{}^{\beta a}+\eta^{\beta\gamma}x_{[\gamma}\Sigma_{\alpha]}{}^a,
\end{equation}
and $x^\beta$ are the Minkowski coordinates corresponding to the tetrad field $e_\beta{}^a$.
The above two equations present the Lorentz-covariant EM and AM currents in Lorentz SR.

On account of Eqs. (\ref{J}), (\ref{sigma}) and (\ref{tau}),
\begin{equation}\label{J2}
J^a=A^\alpha{}_\beta\tau_\alpha{}^{\beta a}+v^b\Sigma_b{}^a,
\end{equation}
and hence
\begin{equation}\label{dJ2}
\partial_aJ^a=\Sigma_\alpha{}^\beta(\partial_\beta v^\alpha-A^\alpha{}_\beta),
\end{equation}
where $\partial_\beta=\partial/\partial x^\beta$.
If the diffeomorphisms $\phi_t$ are isometric transformations
entangled with the Lorentz rotations $g_t$, or explicitly,
\begin{equation}\label{vSR}
v^a=A^\alpha{}_\beta x^\beta e_\alpha{}^a+B^\alpha e_\alpha{}^a,
\end{equation}
where $B^\alpha$ are constants, then a conservation law $\partial_a J^a=0$
follows, where $J^a$ given by Eq. (\ref{J2}) becomes
\begin{eqnarray}\label{JSR}
J^a&=&A^\alpha{}_\beta(\tau_\alpha{}^{\beta a}+x^\beta\Sigma_\alpha{}^a)
+B^\alpha \Sigma_\alpha{}^a\nonumber\\
&=&A^\alpha{}_\beta J_\alpha{}^{\beta a}+B^\alpha J_\alpha{}^a,
\end{eqnarray}
which is the Lorentz-invariant conserved current
with respect to the one-parameter group of symmetry transformations given by
Eqs. (\ref{symtran}) and (\ref{vSR}).

Note that the generalizations of $J_\alpha{}^a$ and $J_\alpha{}^{\beta a}$ in
Lorentz gravity \cite{Kibble,Hehl76} are variational derivatives with respect to the tetrad field
and Lorentz connection. For the coupling system of matter and gravity, these
variational derivatives should be equal to zero due to the gravitational
equations. On the other hand, the Lorentz-invariant $J^a$ can be generalized to
a nonzero current for the coupling
system, which is conserved in the sense that its torsion-free divergence vanishes
\cite{Obukhov06,Obukhov08,Lu15}.

\section{de Sitter special relativity}
Consider a dS-covariant matter field $\psi$ on a dS space, with the action integral and Lagrangian
function as follows:
\begin{equation}\label{LdS}
S=\int_\Omega\mathscr{L}\epsilon,\ \
\mathscr{L}=\mathscr{L}(\psi,d_a\psi,c.c.,\xi^A,d_a\xi^A),
\end{equation}
where $d_a$ is the ordinary exterior derivative, $\xi^A$ are the
5d Minkowski coordinates restricted to the 4d dS hypersurface ${\cal M}_l$
defined by $\eta_{AB}\xi^A\xi^B=l^2$, $\eta_{AB}={\rm diag}(-1,1,\cdots 1)$,
$A,B=0,1,\cdots 4$ are dS indices, and $l$ is a constant with the
dimension of length.
Note that the dS metric $g_{ab}=\eta_{AB}(d_a\xi^A)(d_b\xi^B)$ enters
the Lagrangian via $d_a\xi^A$. Suppose that
\begin{equation}\label{symtrandS}
\Omega\rightarrow\Omega_t=\phi_t[\Omega],\ \
\psi\rightarrow \psi_t=T(g_t)\phi_{t\ast}\psi,\ \
\xi^A\rightarrow \xi_t^A=g_t^A{}_B\phi_{t\ast}\xi^B
\end{equation}
are symmetry transformations forming a one-parameter group, where $g_t=g_t^A{}_B$ are elements
of the special dS group $SO(1,4)$. Provided the matter field equation is satisfied,
the invariance of the action integral under Eq. (\ref{symtrandS}) leads to
\begin{equation}
\mathring{\nabla}_aJ^a+\frac{\partial\mathscr{L}}{\partial\xi^A}\delta\xi^A
+\frac{\partial\mathscr{L}}{\partial d_a\xi^A}\delta(d_a\xi^A)=
\mathscr{L}\mathring{\nabla}_av^a,
\end{equation}
where $\mathring{\nabla}_a$ is the metric-compatible, torsion-free covariant derivative, and
\begin{equation}\label{JdS}
J^a=\frac{\partial\mathscr{L}}{\partial d_a\psi}\delta\psi+c.c.
+\mathscr{L}v^a.
\end{equation}
The arbitrariness of $A^A{}_B\equiv\delta g^A{}_B$, $v^a$ and $\mathring{\nabla}_av^b$
at any given point results in
\begin{equation}\label{dsigma}
\mathring{\nabla}_a\Sigma_b{}^a=\frac{\partial\mathscr{L}}{\partial\xi^A}d_b\xi^A
+\frac{\partial\mathscr{L}}{\partial d_a\xi^A}\mathring{\nabla}_b\mathring{\nabla}_a\xi^A,
\end{equation}
\begin{equation}\label{dtau}
\mathring{\nabla}_a\tau_{AB}{}^a=-\left(\frac{\partial\mathscr{L}}{\partial\xi^{[A}}\xi_{B]}
+\frac{\partial\mathscr{L}}{\partial d_a\xi^{[A}}d_a\xi_{B]}\right),
\end{equation}
\begin{equation}
-\frac{\partial\mathscr{L}}{\partial d_a\psi}d_b\psi+c.c.
=\frac{\partial\mathscr{L}}{\partial d_a\xi^A}d_b\xi^A,
\end{equation}
where
\begin{equation}\label{sigmadS}
\Sigma_b{}^a=-\frac{\partial\mathscr{L}}{\partial d_a\psi}d_b\psi+c.c.
+\mathscr{L}\delta^a{}_b
\end{equation}
is the orbital EM tensor,
\begin{equation}\label{taudS}
\tau_A{}^{Ba}=\frac{\partial\mathscr{L}}{\partial d_a\psi}T_A{}^B\psi+c.c.
\end{equation}
is the dS-covariant spin current, $T_A{}^B$ are representations of the dS
generators, and the dS indices are lowered by $\eta_{AB}$. Let us define
\begin{equation}\label{sigmaAB}
\Sigma_A{}^{Ba}=\Sigma_b{}^a\eta_{AC}(\mathring{\nabla}^b\xi^{[C})\xi^{B]},
\end{equation}
\begin{equation}\label{VABa}
V_A{}^{Ba}=\Sigma_A{}^{Ba}+\tau_A{}^{Ba},
\end{equation}
where the abstract index is raised by the inverse of $g_{ab}$.
With the help of the identities
\begin{equation}
\mathring{\nabla}_a\xi^A\mathring{\nabla}^a\xi^B
=\eta^{AB}-\xi^A\xi^B/l^2,
\end{equation}
\begin{equation}
\mathring{\nabla}_a\mathring{\nabla}_b\xi^A
=-g_{ab}\xi^A/l^2,
\end{equation}
and Eqs. (\ref{dsigma})--(\ref{dtau}), it can be shown that
\begin{equation}\label{dV}
\mathring{\nabla}_aV_A{}^{Ba}=0.
\end{equation}
In order to compare the above equation with Eqs.
(\ref{dsigma=0})--(\ref{dtau=sigma}), let us define
\begin{equation}
V_b{}^a=V_A{}^{Ba}(d_b\xi^A)(2\xi_B/l^2),
\end{equation}
\begin{equation}
V_b{}^{ca}=V_A{}^{Ba}(d_b\xi^A)(d^c\xi_B),
\end{equation}
and $\tau_b{}^a$, $\tau_b{}^{ca}$ in the same way, then it holds that
\begin{equation}\label{tauba}
V_b{}^a=\Sigma_b{}^a+\tau_b{}^a,\ \ V_b{}^{ca}=\tau_b{}^{ca},
\end{equation}
and Eq. (\ref{dV}) leads to
\begin{equation}\label{dVba}
\mathring{\nabla}_aV_b{}^a=\mathring{R}^c{}_{dba}\tau_c{}^{da},
\end{equation}
\begin{equation}\label{dVbca}
\mathring{\nabla}_a\tau_{bc}{}^a=-V_{[bc]},
\end{equation}
where $\mathring{R}^c{}_{dba}$ is the curvature tensor of $\mathring{\nabla}_a$,
and has the expression $\mathring{R}_{cdab}=2g_{a[c}g_{d]b}/l^2$ on ${\cal M}_l$.
A comparison of Eqs. (\ref{dVba})--(\ref{dVbca}) and Eqs. (\ref{dsigma=0})--(\ref{dtau=sigma})
shows that $V_b{}^a$ is the canonical EM tensor in dS SR. In fact, the generalization
of $V_\alpha{}^a$ in dS gravity is just the canonical EM current defined as the variational
derivative with respect to $e^\alpha{}_a$ \cite{Lu15}. According to Eq. (\ref{tauba}), the difference
between $V_b{}^a$ and $\Sigma_b{}^a$ is the inherent EM tensor
\begin{equation}\label{tauba-l}
\tau_b{}^a=\tau_A{}^{Ba}(d_b\xi^A)(2\xi_B/l^2)\sim l^{-1},
\end{equation}
which is originated from the dS spin $\tau_A{}^{Ba}$ and has the order of magnitude of $\Lambda^{1/2}$,
where $\Lambda=3/l^2$ is the cosmological constant. The existence of this inherent EM tensor is an important feature for dS SR. To obtain $\tau_b{}^a$ we need a dS-invariant Lagrangian on
the dS space, but the construction of such a Lagrangian is nontrivial.
Here we give an example for a Dirac field $\psi$ with the Lagrangian
constructed as follows:
\begin{equation}\label{Dirac}
\mathscr L=\frac12{\rm i}(\overline{\psi}\gamma^ad_a\psi-d_a\overline{\psi}\gamma^a\psi)
-{\rm i}m\overline{\psi}\psi,
\end{equation}
where i is the imaginary unit, $\overline{\psi}=\psi^\dagger\gamma^0$,
$\gamma^a=\gamma^Ad^a\xi_A$,
\begin{equation}
\gamma^0={\rm i}\left(
\begin{array}{cc}
0&1\\1&0
\end{array}
\right),\ \
\gamma^i={\rm i}\left(
\begin{array}{cc}
0&\sigma^i\\-\sigma^i&0
\end{array}
\right),
\end{equation}
$i=1,2,3$, $\sigma^i$ are Pauli matrices, and $\gamma^4={\rm i}\gamma^0\gamma^1\gamma^2\gamma^3$.
It is noteworthy that the Lagrangian (\ref{Dirac}) is a function on the dS space, while the
Lagrangian given by Ref. \cite{Gursey} is a function on the 5d Minkowski space. The Dirac-like
equation with respect to Eq. (\ref{Dirac}) is
\begin{equation}
\gamma^ad_a\psi-m\psi-2\gamma^A\xi_A\psi/l^2=0.
\end{equation}
Substitution of Eq. (\ref{Dirac}) and $T^{AB}=\frac14[\gamma^A,\gamma^B]$ into Eqs.
(\ref{taudS}) and (\ref{tauba-l}) yields
\begin{equation}
\tau_b{}^a=\frac{1}{4l^2}{\rm i}\overline{\psi}\gamma^a[\gamma_b,\gamma^A\xi_A]\psi+c.c..
\end{equation}

Now we turn to the problem of how to define the EMA currents in the IC
systems. Without loss of generality, let us discuss the domain on ${\cal M}_l$
with $\xi^4>0$. The inertial-type coordinates on this domain
can be defined as follows \cite{Guo-MPL}:
\begin{equation}
x^\mu=l\xi^\mu/\xi^4,
\end{equation}
where $\mu=0,1,2,3$ is the coordinate index.
Making use of these definitions, $\Sigma_A{}^{Ba}$ defined by Eq. (\ref{sigmaAB})
can be expressed as
\begin{equation}
\Sigma_\mu{}^{4a}=(l/2)M_\mu{}^a,\ \
\Sigma_\mu{}^{\nu a}=\eta^{\nu\sigma}x_{[\sigma}M_{\mu]}{}^a,
\end{equation}
where
\begin{equation}
M_\mu{}^a=\Sigma_\nu{}^a(\delta^\nu{}_\mu+x^\nu x_\mu/l^2),
\end{equation}
$x_\mu=\eta_{\mu\nu}x^\nu$, $\eta_{\mu\nu}={\rm diag}(-1,1,1,1)$, and
$\eta^{\nu\sigma}$ is its inverse. It is seen that $M_\mu{}^a$ can be identified
as the orbital EM current in $\{x^\mu\}$, and $\Sigma_\mu{}^{\nu a}$ can be identified
as the 4d orbital AM current in $\{x^\mu\}$. Then the meaning of $\Sigma_A{}^{Ba}$ becomes clear:
it is a 5d orbital AM current uniting the orbital EMA currents in $\{x^\mu\}$.
Furthermore, $V_A{}^{Ba}$ defined by Eq. (\ref{VABa}) can be expressed as
\begin{equation}
V_\mu{}^{4a}=(l/2)(M_\mu{}^a+S_\mu{}^a),\ \
V_\mu{}^{\nu a}=\eta^{\nu\sigma}x_{[\sigma}M_{\mu]}{}^a+\tau_\mu{}^{\nu a},
\end{equation}
where $S_\mu{}^a=\tau_\mu{}^{4a}(2/l)$, and, $\tau_\mu{}^{4a}$, $\tau_\mu{}^{\nu a}$
are components of $\tau_A{}^{Ba}$. It is seen that $M_\mu{}^a+S_\mu{}^a$ can be identified
as the EM current in $\{x^\mu\}$, and $V_\mu{}^{\nu a}$ can be identified
as the 4d AM current in $\{x^\mu\}$. Then the meaning of the conserved current $V_A{}^{Ba}$ becomes
clear: it is a 5d AM current uniting the EMA currents in $\{x^\mu\}$.
At the coordinate origin, $M_\mu{}^a$ coincides with the orbital EM tensor $\Sigma_b{}^a$,
$S_\mu{}^a$ coincides with the inherent EM tensor $\tau_b{}^a$, and so $M_\mu{}^a+S_\mu{}^a$ coincides
with the canonical EM tensor $V_b{}^a$. Likewise, $V_\mu{}^{\nu a}$ coincides with the spin tensor
$\tau_b{}^{ca}$ at the coordinate origin.

On account of Eqs. (\ref{JdS}), (\ref{sigmadS}) and (\ref{taudS}),
\begin{equation}\label{JdS2}
J^a=A^A{}_B\tau_A{}^{Ba}+v^b\Sigma_b{}^a.
\end{equation}
Provided the diffeomorphisms $\phi_t$ are isometric transformations
entangled with the dS rotations $g_t$, or explicitly,
\begin{equation}\label{vdS}
v^a=(A^A{}_B\xi^B\partial_A)^a,
\end{equation}
where $\partial_A$ are the 5d Minkowski coordinate basis vector fields,
then a conservation law $\mathring{\nabla}_aJ^a=0$ follows, where $J^a$
given by Eq. (\ref{JdS2}) becomes
\begin{equation}\label{JdS3}
J^a=A^A{}_BV_A{}^{Ba},
\end{equation}
which is the dS-invariant conserved AM current with respect to the one-parameter group
of symmetry transformations given by Eqs. (\ref{symtrandS}) and (\ref{vdS}).

Note that the generalization of $V_A{}^{Ba}$ in dS gravity is a variational
derivative with respect to the dS connection \cite{Lu15}. For the coupling system
of matter and gravity, the variational derivative should be equal to zero due to
the gravitational field equation. On the other hand, $J^a$ can be generalized to
a nonzero current for the coupling system, which is conserved in the sense that its
torsion-free divergence vanishes \cite{Lu15}.

\section{Remarks}

The paper presents two kinds of conservation laws for a matter field in dS SR.
The first kind is the conservation of the dS-covariant AM current (\ref{VABa}), which
unites the EMA currents of the matter field in an IC system. The second kind is the
conservation of the dS-invariant AM current (\ref{JdS3}), which can be generalized
to a conserved current for the coupling system of matter and gravity.
Moreover, it is found that the canonical EM tensor contains an inherent
part $\tau_b{}^a\sim\Lambda^{1/2}$ originated from the dS spin $\tau_A{}^{Ba}$.
The existence of the inherent EM tensor $\tau_b{}^a$ is a distinctive phenomenon
in dS SR compared to the ordinary Lorentz SR. It is interesting as a future work to analyze the
observational effect of $\tau_b{}^a$.

\section*{Acknowledgments}
I would like to thank the late Prof. H.-Y. Guo, and
Profs. C.-G. Huang, Z.-B. Li and X.-W. Liu for their help
and the useful discussions. The project is funded by China Postdoctoral Science Foundation
under Grant No. 2015M572393.

\end{document}